# Spin wave propagation in ferrimagnetic Gd$_x$Co$_{1-x}$

Shinsaku Funada[1], Tomoe Nishimura[1], Yoichi Shiota[1*], Shuhei Kasukawa[1], Mio Ishibashi[1], Takahiro Moriyama[1], and Teruo Ono[1,2*]

[1]*Institute for Chemical Research, Kyoto University, Gokasho, Uji, Kyoto 611-0011, Japan*

[2]*Center for Spintronics Research Network, Graduate School of Engineering Science, Osaka University, Toyonaka, Osaka 560-8531, Japan*

*E-mail: shiota-y@scl.kyoto-u.ac.jp, ono@scl.kyoto-u.ac.jp

## Abstract

Recent advances in antiferromagnetic spin dynamics using rare-earth (RE) and transition-metal (TM) ferrimagnets have attracted much interest for spintronic devices with a high speed and density. In this study, the spin wave properties in the magnetostatic backward volume mode and surface mode in RE-TM ferrimagnetic Gd$_x$Co$_{1-x}$ films with various composition *x* are investigated using spin wave spectroscopy. The obtained group velocity and attenuation length are well explained by the ferromagnet-based spin wave theory when the composition of Gd$_x$Co$_{1-x}$ is far from the compensation point.

In magnonics where spin waves are used for information transport and processing,[1–4] yttrium iron garnet ($Y_3Fe_5O_{12}$) and permalloy ($Ni_{80}Fe_{20}$) have been widely used, because they exhibit a low Gilbert damping parameter and soft magnetic properties. However, in such materials, the magnetization precesses in right-handed chirality because they only have one type of magnetic lattice. In contrast, the polarization of spin waves and its manipulation have been theoretically predicted using antiferromagnetic spin waves.[5,6]

The recent progress in the development of antiferromagnetic spintronics has enabled spintronic devices with high speed and density.[7–9] In particular, rare-earth (RE) and transition-metal (TM) ferrimagnets are one of most fascinating materials to investigate antiferromagnetic spin dynamics. As the magnetic moments of TM elements (Fe, Co, Ni, and their alloys) and RE elements (Gd, Tb, etc.) can be antiferromagnetically coupled, the compensation points of the net magnetization $T_M$ or the net angular momentum $T_A$ can be reached by varying the temperature and/or the composition of each element.[10–14] These properties make these materials interesting for investigating phenomena such as ultrafast all-optical switching,[15–17] current-induced switching,[18,19] spin-orbit torque,[20–23] and fast domain wall motion.[24–26] These findings motivated us to study the spin wave properties in ferrimagnetic $Gd_xCo_{1-x}$ alloy with various composition $x$. Note that magnetic properties in YIG show ferromagnetic-like behavior unlike the RE-TM ferrimagnets even though YIG is also a ferrimagnet. This is because that the magnetism in YIG is governed by an uncompensated magnetic moment of $Fe^{3+}$ ions in octahedral and tetrahedral coordinate sites. In addition, we can expect large group velocity in RE-TM ferrimagnets due to large saturation magnetization than that in YIG. Therefore, ferrimagnetic $Gd_xCo_{1-x}$ alloy has a potential to investigate the antiferromagnetic spin waves.

The samples were prepared on thermally oxidized Si substrates using direct current magnetron sputtering. The film stack consisted of $Gd_xCo_{1-x}$ (20 nm)/Pt (2 nm)/Ta (5 nm), wherein the Pt/Ta bilayer was used as the capping layer. Because Gd is a very reactive element and easily oxidized, we selected Pt/Ta bilayer for capping layer rather than an oxide capping layer, such as $Al_2O_3$ or $SiO_2$. The enhancement of the Gilbert damping originated from the spin pumping effect is not negligible, even though it should be small due to the large thickness of $Gd_xCo_{1-x}$.[27,28] Therefore, the estimated magnetic properties of $Gd_xCo_{1-x}$ in the following experiments include the effect of metal capping layer. However, it should be emphasized that the metal capping layer does not affect discussions on the composition dependence of magnetic properties in $Gd_xCo_{1-x}$. The $Gd_xCo_{1-x}$ alloy was deposited using the co-sputtering method from Gd and Co targets with different sputtering powers. The compositions of the Gd and Co atoms were calculated from the deposition rates, which were calibrated in advance with X-ray reflectivity. We selected $Gd_xCo_{1-x}$ compositions with $x =$ 0.22, 0.30, 0.40, and 0.59 to investigate the spin wave propagation in both Co-rich and Gd-rich $Gd_xCo_{1-x}$ alloys. All the samples exhibited an in-plane magnetic easy axis.

First, the static magnetic properties were measured using superconducting quantum interference device magnetometry. Figure 1(a) shows the saturation magnetization $M_s$ as a function of the temperature for the samples with $x =$ 0.22, 0.30, 0.40, and 0.59. In the temperature range of 20–300 K, the magnetic moment of Co (Gd) was dominant for the samples with $x =$ 0.22, 0.30 ($x =$ 0.59). We found the magnetization compensation temperature $T_M$ to be approximately 170 K for the sample with $x =$ 0.40, which deviates from the previous reported compensation point of around $x =$ 0.2.[12] This might be due to the deviation between expected and actual composition of $Gd_xCo_{1-x}$ and/or the thickness dependence of the compensation point in RE-TM ferrimagnets.[14] Then, the magnetization curves under perpendicular magnetic field were investigated for the samples with $x =$ 0.22, 0.30, and 0.59, as shown in Figs. 1(b)-(d). All the samples exhibit an in-plane magnetic easy axis. The effective saturation magnetization $M_{s,eff} = M_s - H_\perp$, where $H_\perp$ is the perpendicular magnetic anisotropy field, was estimated from the perpendicular saturation magnetic field. Table I lists the obtained values of $M_s$ (300 K) and $M_{s,eff}$. The reduction in $M_{s,eff}$ compared with $M_s$ (300 K) results from the perpendicular magnetic anisotropy, which originates from the bulk perpendicular magnetic anisotropy.[29,30]

Next, the spin wave spectra were measured using a vector network analyzer (VNA) at room temperature. Figure 2 shows the top view of the spin wave device. The films were structured into spin wave waveguides of sizes 50 × 100 μm² ($x =$ 0.22, 0.30, and 0.40) and 100 × 100 μm² ($x =$ 0.59) using electron-beam lithography and Ar ion milling. After depositing an 80 nm-thick $SiO_2$ layer for electrical isolation, shortened coplanar waveguides (CPWs) were patterned on top of the spin wave waveguides using electron-beam lithography and lift-off process of sputter-deposited Ti (5 nm)/Au (100 nm). Two CPWs were designed on a signal line (2 μm) and two ground lines (1 μm) with a gap of 1 μm, which effectively excited or detected the spin waves with a wavenumber of 1.2 μm⁻¹, as shown in Fig. 2(b).[31] The injected microwave power of the VNA was −5 dBm, which was low enough to maintain the linear response region of magnetization dynamics. Depending on the direction of the applied magnetic field, the spin waves in the magnetostatic backward volume wave (MSBVW) and magnetostatic surface wave (MSSW) configurations were excited,[32] as shown in Fig. 2. From the self-scattering parameters $S_{11}$ and $S_{22}$ and the mutual-scattering parameters $S_{12}$ and $S_{21}$, we extracted the local spin wave resonance under the CPWs and the propagation characteristics of the spin waves between the two CPWs, respectively. Note that we could not observe the resonance peaks for the sample with $x = 0.40$ because of the low net saturation magnetization. Hereafter, we discuss the spin wave properties of the samples with $x = 0.22$, 0.30, and 0.59.

We measured the local spin wave resonance under the CPWs, not the propagating spin wave resonance, by means of $S_{11}$ measurements in the MSBVW configuration, where the magnetic field was applied along the spin wave propagation direction. To suppress the frequency-dependent background, including circuit resonances and losses, we measured the spectra by sweeping the magnetic field. Figure 3(a) shows the Re[$S_{11}$] spectra in the MSBVW configuration under various frequencies for the sample with $x = 0.22$. Figures 3(b) and 3(c) show the resonance frequency and linewidth. Given the small dispersion in the MSBVW configuration in the limit of $kt \ll 1$, where $k$ and $t$ denote the wavenumber of the spin waves and the thickness of the $Gd_xCo_{1-x}$ layers, respectively, the resonance frequency $f_{MSBVW}$ and full width at half maximum of the resonance peak $\Delta H$ can be expressed as follows.[33]

$$f_{MSBVW} = \frac{\mu_0 \gamma}{2\pi}\sqrt{H\left(H + M_{s,eff}\frac{1-e^{-kt}}{kt}\right)} \simeq \frac{\mu_0 \gamma}{2\pi}\sqrt{H(H + M_{s,eff})} \quad (1)$$

$$\Delta H = \Delta H_0 + \frac{4\pi\alpha}{\mu_0 \gamma} f_{MSBVW} \quad (2)$$

where $\gamma = (g\mu_B)/\hbar$ is the gyromagnetic ratio, $g$ is the Landé g-factor, $\mu_B$ is the Bohr magneton, $\hbar$ is the Dirac's constant, $\mu_0$ is the permeability of free space, $H$ is the applied magnetic field, $\Delta H_0$ is the inhomogeneous linewidth broadening, and $\alpha$ is the Gilbert damping parameter. From the fittings obtained using Eqs. (1) and (2), $g$, $M_{s,eff}$, and $\alpha$ were estimated for the samples with various compositions, as summarized in Table II. $M_{s,eff}$ obtained from $M$-$H$ curves and MSBVW measurements are more or less consistent. The $g$ in Co-rich sample ($x = 0.22$) shows higher value than that in Gd-rich sample ($x = 0.59$), and $g$ and $\alpha$ takes maximum for the sample with $x = 0.30$, which is near the angular momentum compensation point in the range of composition that we examined. This compositional dependence of $g$ and $\alpha$ can be explained with the simple mean field model,[11,34] where $g\mu_B/\hbar = M_{net}/A_{net}$ and $\alpha = A_0/A_{net}$. Theoretically these two values diverge at the vicinity of the angular momentum compensation point, and the tendencies we observed in this study are consistent. M. Binder et al.[12] previously reported the composition dependence of $g$ and $\alpha$ in ferrimagnetic GdCo, which has a large difference from our results. Because they only examined $x = 0.2 - 0.25$ composition range of GdCo, which is the vicinity of the angular momentum compensation point, simple comparisons cannot be made. However we find that our results are comparable to the reported values in other RE-TM ferrimagnetic systems, e.g. GdFeCo.[11,13]

The propagating spin waves in the MSSW configuration, where the magnetic field was applied perpendicular to the spin wave propagation direction, were investigated. Figures 4(a)–(c) show the Re[$S_{21}$] and Im[$S_{21}$] spectra for samples with $x$ = 0.22, 0.30, and 0.59, respectively, under a magnetic field of 10 mT. The distance $d$ between the two CPWs for each sample is indicated in Fig. 4. The oscillating signatures indicate the spin wave propagation between the two CPWs.[33] We estimated the spin wave group velocity $v_g$ using the relationship $v_g = \Delta f \cdot d$, where $\Delta f/2$ corresponds to the frequency difference between the positive and negative peaks in the Im[$S_{21}$] spectra, as shown in Fig. 4(b). The insets in Figs. 4(a) and (c) show the intensity of the transmittance signal $|S_{21}|$ as a function of $d$. The fittings of the exponential decay function yield the spin wave attenuation length $L_{att}$. Because of the low signal intensity and high damping constant, we could not evaluate $L_{att}$ from $d$ dependence of $|S_{21}|$ for the sample with $x$ = 0.30.

From the theoretical dispersion relationship in the MSSW configuration, $f_{MSSW} = \frac{\mu_0 \gamma}{2\pi} \sqrt{H(H + M_{s,eff}) + \frac{M_s M_{s,eff}}{4}(1 - e^{-2kt})}$, $v_g$ and $L_{att}$ can be calculated as follows,

$$v_g = \frac{(\mu_0 \gamma)^2 M_{s,eff} \cdot M_s d}{8\pi f} \quad (3)$$

$$L_{att} = \frac{v_g}{\mu_0 \gamma (2H + M_{s,eff}) \alpha} \quad (4)$$

where, $v_g = 2\pi \, \partial f_{MSSW}/\partial k$ and $(\mu_0 \gamma (2H + M_{s,eff}) \alpha)^{-1}$ represent life-time of the magnetization precession. Table II lists the experimentally and theoretically obtained $v_g$ and $L_{att}$ values. The theoretically calculated $v_g$ and $L_{att}$ qualitatively reproduce the experimental results. The overestimation of $v_g$ in the experiments may have resulted from the finite widths of the CPWs.[35] According to Table II, the spin wave observed in this study can be explained by the typical magnetostatic spin wave theory based on ferromagnets.

In conclusion, we investigated the spin wave properties in RE-TM ferrimagnetic $Gd_xCo_{1-x}$ films with various composition $x$. In the MSBVW configuration, $g$, $M_{s,eff}$, and $\alpha$ were estimated. In the MSSW configuration, the obtained $v_g$ and $L_{att}$ were qualitatively consistent with the ferromagnet-based spin wave theory when the composition of $Gd_xCo_{1-x}$ was far from the compensation point.

**Acknowledgments**

This work was partially supported by JSPS KAKENHI (Grant Numbers 26103001, 15H05702, 16H05977, 18K19021).

**Figure Captions**

**Fig. 1.** (a) Temperature dependence of magnetization under in-plane magnetic field of 100 mT for Gd$_x$Co$_{1-x}$ with various values of $x$. Right figures show the magnetic moments of sublattice, Co (blue) and Gd (yellow), with various composition at room temperature. The dominated element of the net magnetization transitions between Co and Gd depending on the composition $x$. (b)-(d) Perpedicular magnetization curves for the sample with (b) $x = 0.22$, (c) $x = 0.30$, and (d) $x = 0.59$. Red lines represent the linear fittings of the magnetization in below and above saturation magnetic field, and $M_{s,\text{eff}}$ for each sample was extracted from the intersection of two linear fittings.

**Fig. 2.** (a) Top view of a spin wave device. The distance between the two CPWs is defined as $d$. Depending on the direction of the magnetic field, spin waves with magnetostatic backward volume wave (MSBVW) and magnetostatic surface wave (MSSW) configurations are investigated. (b) Fourier transform of the current distribution in the designed CPWs.

**Fig. 3.** (a) Frequency dependence of Re[$S_{11}$] spectra under external magnetic field in the MSBVW configuration for Gd$_{0.22}$Co$_{0.78}$. (b) Resonance frequency as a function of the magnetic field, and (c) full width at half maximum of the resonance peak as a function of the frequency for Gd$_x$Co$_{1-x}$ with various $x$ values. The solid lines in (b) and (c) indicate the fittings obtained using Eqs. (1) and (2), respectively.

**Fig. 4.** (a)–(c) Propagating spin wave spectra under an external magnetic field of 10 mT for (a) Gd$_{0.22}$Co$_{0.78}$ with $d = 10$ μm, (b) Gd$_{0.30}$Co$_{0.70}$ with $d = 8$ μm, and (c) Gd$_{0.59}$Co$_{0.41}$ with $d = 10$ μm. The insets in (a) and (c) show the intensity of the transmittance signal $|S_{21}|$ as a function of $d$.

**Table I.** Experimentally determined $M_s$ (300 K) and $M_{s,eff}$ of $Gd_xCo_{1-x}$ measured from SQUID measurements.

| $x$ | $M_s$ (300 K) [MA/m] | $M_{s,eff}$ [MA/m] |
|---|---|---|
| 0.22 | 1.06 | 0.66 |
| 0.30 | 0.67 | 0.49 |
| 0.59 | 0.55 | 0.52 |

**Table II.** Experimentally determined magnetic properties of $Gd_xCo_{1-x}$ measured from spin wave measurements. The values in parentheses under $v_g$ and $L_{att}$ were theoretically calculated using Eqs. (3) and (4), respectively.

| $x$ | $M_{s,eff}$ [MA/m] | $g$ | $\alpha$ | $v_g$ [km/s] | $L_{att}$ [μm] |
|---|---|---|---|---|---|
| 0.22 | 0.68 | 2.26 | 0.016 | 10.4 (7.5) | 2.13 (2.60) |
| 0.30 | 0.32 | 2.68 | 0.032 | 7.5 (3.2) | (0.98) |
| 0.59 | 0.45 | 1.90 | 0.016 | 3.2 (2.8) | 1.65 (1.85) |

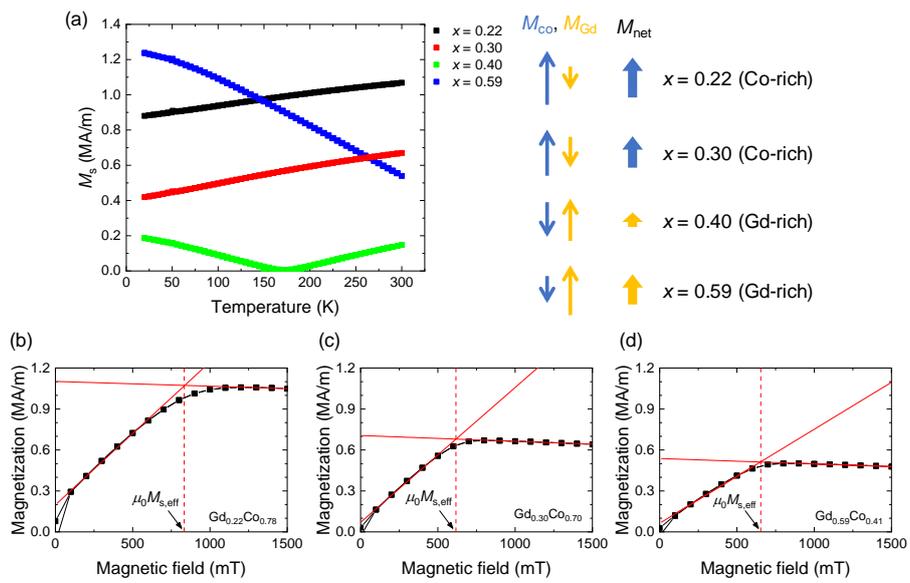

Fig. 1

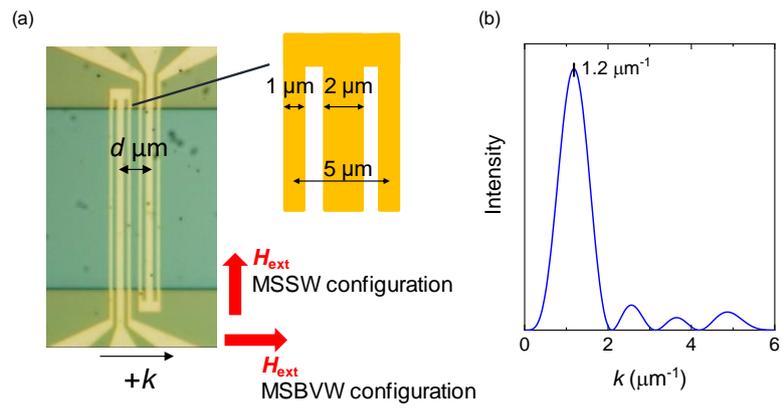

Fig. 2

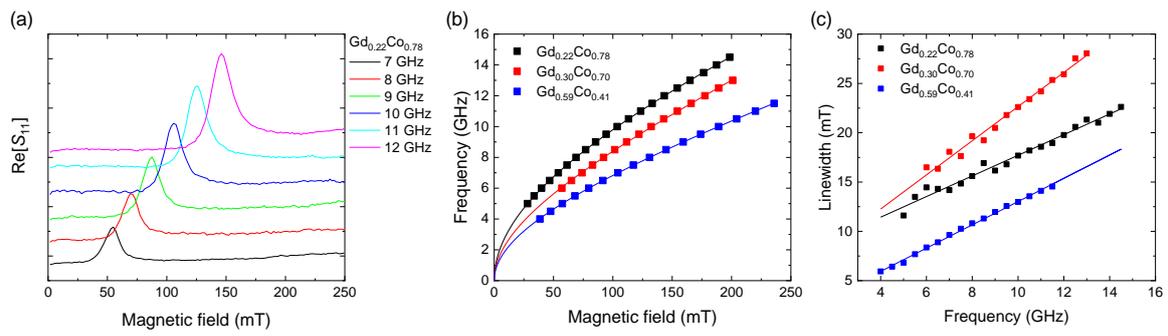

Fig. 3

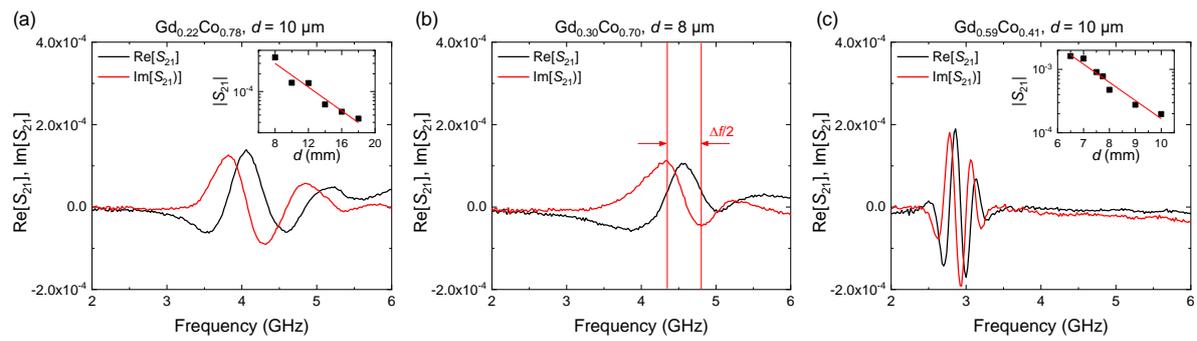

Fig. 4